\documentclass{elsart}

\usepackage{graphicx}
\usepackage{amssymb}

\def\NIMA#1#2#3{Nucl. Instr. Meth. Phys. Res. A {\bf #1}\ (#2)\ #3}
\def\NIM#1#2#3{Nucl. Instr. Meth. {\bf #1}\ (#2)\ #3}

\begin{document}

\begin{frontmatter}

\title{Drift velocity and gain in argon- and xenon-based mixtures}

%

\author[gsi]{A.~Andronic},
\author[liverpool]{S.~Biagi},
\author[gsi]{P.~Braun-Munzinger}, 
\author[gsi]{C.~Garabatos}, 
\author[gsi]{and G.~Tsiledakis} 

\address[gsi]{Gesellschaft f{\"u}r Schwerionenforschung, Darmstadt, Germany}
\address[liverpool]{Department of Physics, University of Liverpool, UK}


\begin{abstract}
We present measurements of drift velocities and gains in gas mixtures based on
Ar and Xe, with CO$_2$, CH$_4$, and N$_2$ as quenchers, 
and compare them with calculations. In
particular, we show the dependence of Ar- and Xe-CO$_2$ drift velocities and
gains on the amount of nitrogen contamination in the gas, which in real
experiments may build up through leaks. A quantification of the Penning 
mechanism which contributes to the Townsend coefficients of a given gas mixture
is proposed.

\end{abstract}

\end{frontmatter}

\section{Introduction} \label{aa:intro}
Modern detectors such as those being 
constructed for the Large Hadron Collider (LHC) include large-volume 
gaseous detectors which
are expected to operate continuously for several months every year.
The required performance of these detectors usually involves the
precise control and monitoring of the drift velocity and the gain. 
These relevant gas parameters depend on the detector field configuration 
and on the gas components, composition, density, and purity.
The gas mixtures used in these detectors, driven by performance,
may contain a high cost component which makes it important to recirculate
the gas in a closed loop, with a modest rate of fresh gas injection.
This mode of operation results, if no gas envelope is foreseen around the
detector, in an increasing amount of air in the
gas volume, entering through leaks, however small these are.
While oxygen can be readily removed by appropriate filters,
the known methods for nitrogen removal are complex and tedious, 
and lead to further losses of the main gas and to the modification of
its composition. The latter circumstance is particularly undesirable during
data taking periods. Thus, nitrogen gradually builds up into the mixture.
The operation of the detector, namely the charge transport
and amplification processes, may therefore change during running periods.

A case in point is the ALICE Transition Radiation Detector 
(TRD) \cite{aa:tdr}, which is designed
to provide electron identification and particle tracking in the
high-multiplicity environment of Pb+Pb collisions at the LHC.
To achieve these goals, accurate pulse height 
measurements over the full drift time (order of 2~$\mu$s) in wire chambers 
operated with a Xe-CO$_2$~[85-15] gas is necessary.
Therefore, knowledge of both the drift velocity and the gas gain is 
important.
The large volume (28~m$^3$) of this barrel detector and the high cost of xenon 
makes it mandatory to recirculate the gas mixture,
and to limit the injection of fresh gas to replenish what is lost
through leaks only. In this case, the nitrogen
concentration in the mixture would reach 8~\% after a running period of
8 months. Most of this nitrogen may be cryogenically distilled 
and removed from the mixture during the shutdown periods, 
at a moderate loss of xenon. Since the gas mixture during running
periods contains a varying admixture of N$_2$, studies of the influence
of N$_2$ on the gas properties become very important.

In the following section we describe the experimental method used to measure
both the drift velocity and the gain of various gas mixtures.
In section~\ref{aa:vd} we show the measured drift velocities. 
Measurements of drift velocities in some binary and ternary Xe-based 
mixtures (without nitrogen)
have been published earlier \cite{aa:chris,aa:dolg,aa:kunst,aa:becker}.
We compare our results to existing data where available and to 
simulations, in order to validate our method. 
The gain measurements, together with results from 
simulations, are presented in section~\ref{aa:n2g}. 
We finally draw our conclusions.

\section{Experimental setup} \label{aa:meth} 

The experimental results are obtained using a small drift 
chamber with 
a geometry similar to that anticipated for the final ALICE TRD
\cite{aa:tdr}, but with a much smaller active area (10$\times$10~cm$^2$).
The chamber has a drift region of 31.5~mm and an amplification region of 
6.4~mm.
The anode wires (W-Au, 20~$\mu$m diameter) have a pitch of 5~mm.
For the cathode wires (Cu-Be, 75~$\mu$m diameter) the pitch is 2.5~mm.
The signal is read out on a cathode plane segmented into rectangular pads of 
area 6~cm$^2$. 
The drift electrode is made of a 25~$\mu$m aluminized Kapton foil, 
which also serves as gas barrier. The electric
 field thus created is sufficiently uniform over the full active area of the 
pad plane. 

\begin{figure}[hbt]
\centering\includegraphics[scale=0.45]{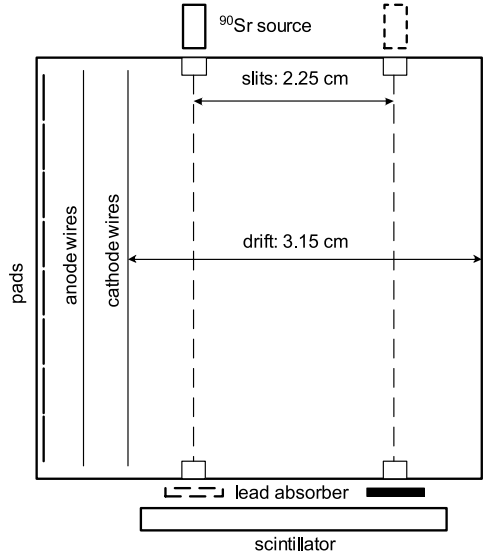}
\caption{Schematics of the modified drift chamber used for the 
drift velocity measurements.}
\label{aa:prin} 
\end{figure} 

A schematic view of the setup used for the drift velocity 
measurements is presented in Fig.~\ref{aa:prin}.
Two sets of slits, 
0.75~mm wide, are machined into the stesalit side walls of the drift region 
and covered with Kapton foils. 
Electrons from a collimated $^{90}$Sr
radioactive source enter the drift volume through either of these slits, and
ionise the gas. Some of these particles cross the corresponding 
slit at the other side of the drift enclosure, behind which a 
scintillator is placed for triggering purposes. Triggered events will 
show signals in the 8 pads under the track, with a drift 
time corresponding, on average, to the distance from the selected set of 
slits to the anode plane. A 2~mm thick lead absorber is placed behind the 
opposite outlet slit to prevent triggers from particles going at an angle
through the detector.
For each set of slits, we record on FADCs the pulse height distributions 
on the pads as a function of the drift time of the tracks.
The corresponding average times are evaluated and then subtracted. 
In this way, the contribution to the drift time
of the amplification region, where the electric field is not uniform, is
cancelled. The anode voltage is adjusted for each mixture to achieve a 
gain near 10$^4$, and ranges between 1450~V and 1800~V.
Both the pad plane and the cathode wires are kept at ground potential.
The amplification field leaks through the cathode wire plane 
and effectively increases the drift
field. In order to correct for this effect, the position of the 0~Volts
equipotential line, relative to the position of the cathode wires,
 is computed with the Garfield simulation package \cite{aa:garf} for each
set of anode and drift voltages. This shift, which depends on both the
drift and anode voltages, ranges in our case from 0.02~mm to 6~mm. 
The reduced electric field is finally evaluated with the recorded
ambient pressure.
The oxygen and water vapour in the gas was monitored during the measurements,
and varied, depending on the gas flow, between 10 and 50~ppm O$_2$, and
300 to 500~ppm H$_2$O.
The resulting drift velocity, measured as a function of the 
reduced electric field, has an uncertainty estimated to be lower than 10~\%.

The gain is measured with an $^{55}$Fe source, by counting the number 
of signals produced by X-rays
absorbed in the gas, and measuring the currents drawn by the anode high 
voltage power supply due to these photons. Typical rates are 60~kHz in a 
projected area of order 1~cm$^2$.  The number of
primary electrons per photon produced in the gas is derived for each 
mixture separately using the work functions given in \cite{aa:sauli}.
The drift voltage during these measurements was set at -2~kV.

We use a prototype of the charge-sensitive preamplifier/shaper (PASA)
especially designed and built for TRD prototypes with discrete components.
It has a noise on-detector of about 2000 electrons r.m.s. and the FWHM
of the output pulse is about 120~ns for an input step function. 
The nominal gain of the PASA is 3~mV/fC.
The FADC has an 8-bit non-linear conversion 
and adjustable baseline, and runs at 100~MHz sampling frequency.

\section{Drift velocity measurements} \label{aa:vd} 
\begin{figure}[hbt]
\centering\includegraphics[width=.63\textwidth]{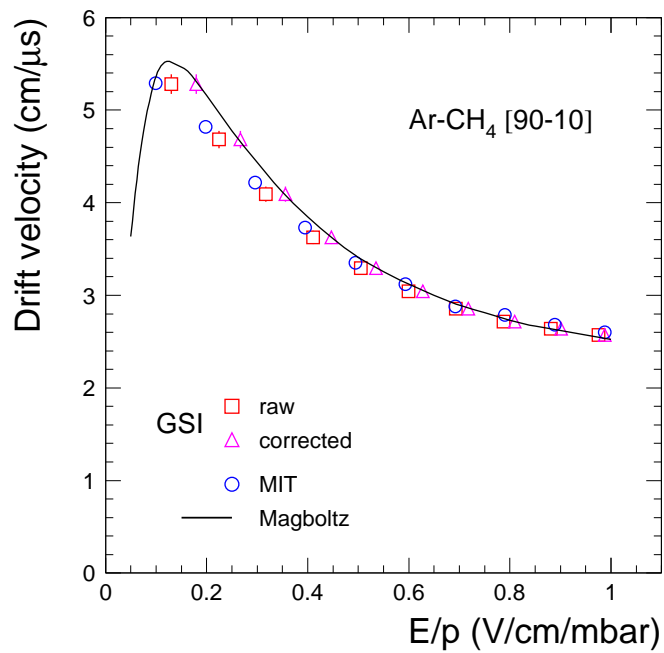}
\caption{Drift velocity measurements and calculations in Ar-CH$_4$~[90-10].
The effect of the anode potential on the configuration of the electric drift
field manifests itself (square data points) especially at low fields,
and is corrected for (triangles).}
\label{aa:arch4} 
\end{figure}

In order to validate our experimental method, we first measured the drift 
velocity of a well known mixture, Ar-CH$_4$~[90-10], and compared our results 
with existing data \cite{aa:becker} that we refer to as MIT data. 
We also compare the measurements with Magboltz \cite{aa:magb} 
calculations. We show our results without and with the correction of
the reduced field due to the leakage of the anode field into
the drift region. This correction is higher at lower drift fields.
As can be seen in Fig.~\ref{aa:arch4}, the agreement between this work 
and the calculation is good only after the correction of the drift 
field values.
On the other hand, a clear discrepancy with the MIT data is visible 
at low fields, and reaches 10~\%. 

\begin{figure}[hbt]
\centering\includegraphics[width=.65\textwidth]{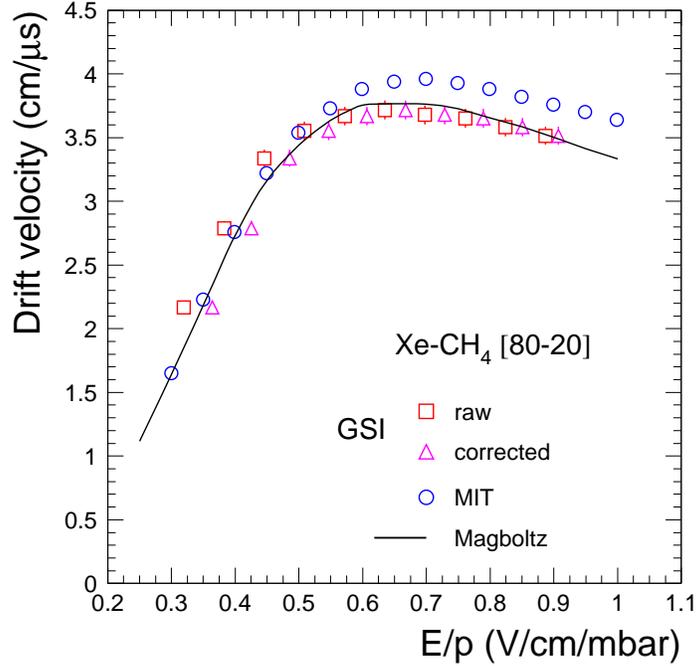}
\caption{Drift velocity in Xe-CH$_4$~[80-20], with (triangles) and 
withough (squares) drift field correction as measured in this work, 
together with other measurements and a calculation.}
\label{aa:xech4} 
\end{figure} 

We then measured the drift velocity of Xe-CH$_4$~[80-20], 
which the MIT group has also measured.
The motivation for this second reference measurement is that the multiple 
scattering of electrons coming for the $^{90}$Sr source is significant in 
xenon. This effect, in combination with the asymmetric gas volume 
available for tracks emerging from either slit, biases the measurement
towards larger drift velocities by, in this case, as much as 15~\%.
For this reason, we work with drift time distributions measured 
on the pad closest to the entrance slit only, where multiple scattering 
is minimal. The argon data showed no 
difference, within 2~\%, in the results obtained from any pad, 
meaning that the drift field
is uniform enough in the regions above the pads at the edges of the active 
area.
The resulting drift velocity and its comparisons are shown in 
Fig.~\ref{aa:xech4}.
There is again a significant discrepancy between our measurement and 
the MIT results at fields above drift velocity saturation. 
However, the calculations of the drift velocity in this region 
are compatible with the measurement of this work.
At low fields, on the contrary, the MIT data agree well with the 
calculation, whereas our results underestimate the calculated values by
7~\% near 0.45~V/cm/mbar.

\begin{figure}[hbt]
\centering\includegraphics[width=.65\textwidth]{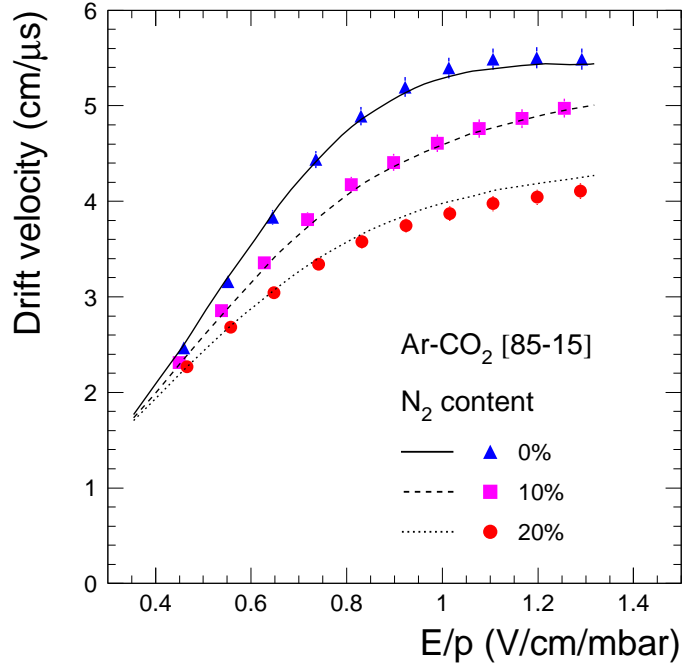}
\caption{Drift velocity in Ar-CO$_2$~[85-15] with N$_2$ additions and
comparison with simulations.}
\label{aa:arco2} 
\end{figure} 

\begin{figure}[hbt]
\centering\includegraphics[width=.65\textwidth]{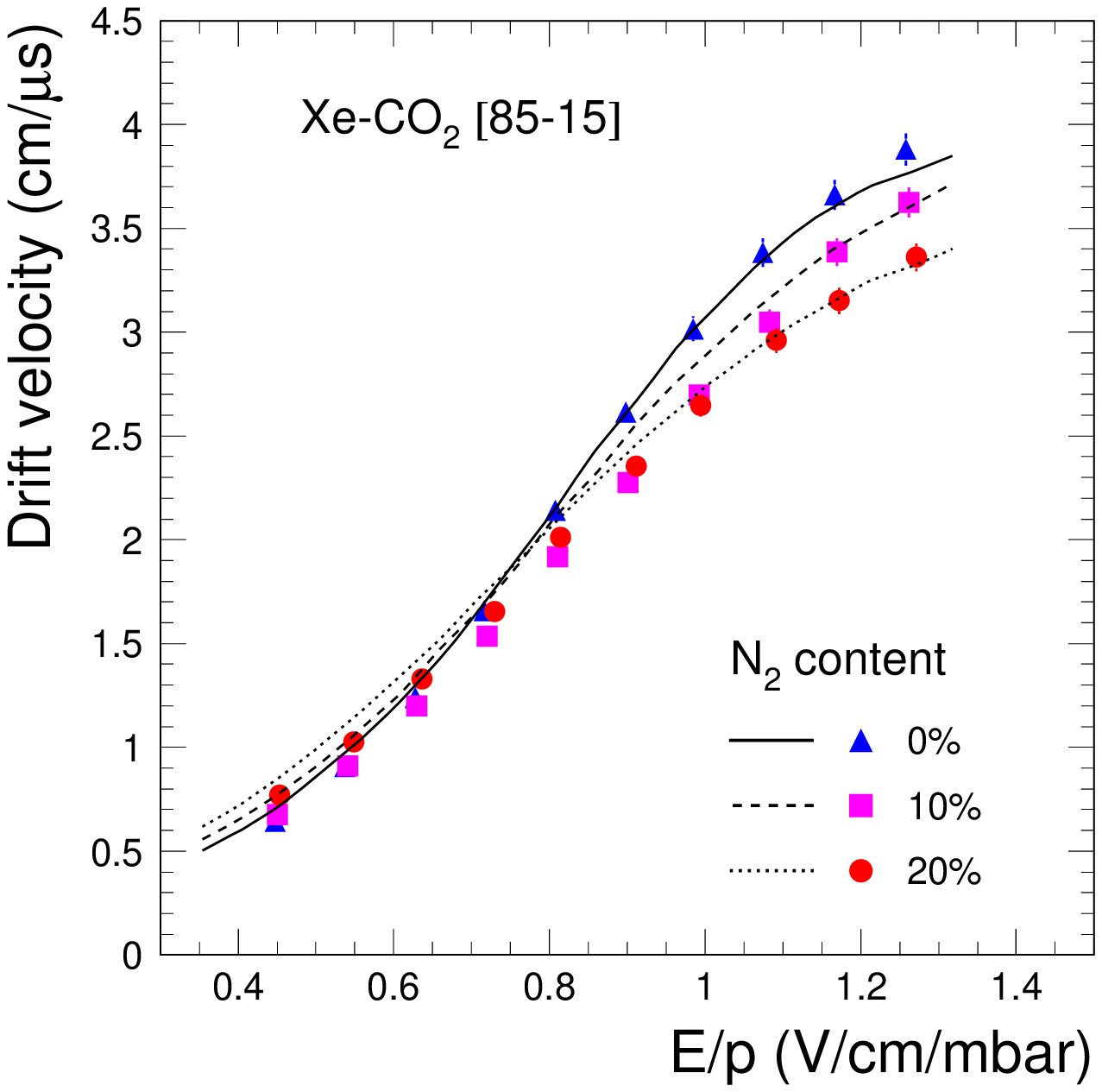}
\caption{Drift velocity in Xe-CO$_2$~[85-15] with N$_2$ additions.}
\label{aa:xeco2} 
\end{figure} 

The next set of measurements were undertaken for Ar-CO$_2$ [85-15] and 
admixtures of 0, 10 and 20~\% N$_2$. Adding, for example, 10~\% N$_2$ 
into the mixture results in an Ar-CO$_2$-N$_2$ [76.5-13.5-10] mixture.
As can be seen in Fig.~\ref{aa:arco2}, the 
drift velocity decreases with increasing concentration of nitrogen, 
and there is a reasonable agreement between measurements
and simulation. Due to the saturation of the drift velocity at lower
values with increasing N$_2$ content, keeping the drift velocity constant
would require higher and higher drift voltages as the gas composition changes,
and to maintain a fast mixture would eventually become impossible.

Finally, the results for Xe-CO$_2$ [85-15] mixtures with 0, 10 and 20~\% N$_2$ 
admixtures, shown in Fig.~\ref{aa:xeco2}, exhibit a weak dependence on
the nitrogen concentration.
We notice deviations of up to 12~\% with respect to the calculations at
intermediate fields. The calculated drift velocities 
exhibit a crossing of the three curves at a field near 800~V/cm.
The measurements show very little dependence of the drift velocity on the
N$_2$ concentration at fields up to this value.
Since, for example, the anticipated electric field of the ALICE TRD 
is 700~V/cm, this circumstance should be welcome: the change in drift 
velocity due to substantial accumulations of nitrogen would be negligible 
in this case.

\section{Gain measurements} \label{aa:n2g} 
As explained in the introduction, the absolute gain as a function of
the anode voltage is measured with the use of a $^{55}$Fe source, which is
placed in front of the entrance window of the chamber.
We have also carried out calculations of the gain with the use of the package 
Magboltz 2 \cite{aa:magb}. 
This program computes the Townsend coefficient for a given
gas mixture and electric field. By introducing this information,
 together with the chamber geometry and voltages, into Garfield 
\cite{aa:garf}, one can
calculate the gain of the detector for each mixture and anode voltage.
The multiplication factor obtained this way accounts for the electrons
produced in the avalanche by collisions of atoms or molecules with 
other energetic electrons.
In addition, Magboltz 2 provides information about the excited and 
ionised species produced in the avalanche. This information can be 
used to scale up the Townsend coefficients, according to the
ionisation of gas species due to collisions with other excited metastable
gas states (Penning effect) \cite{aa:mill,aa:velaz,aa:cwet}. 
Since this energy transfer rate is a priori not known, 
the experimental data are used as a guide to tune one parameter, 
the so-called Penning fraction, for matching the calculations to the 
measurements. The Penning fraction refers to the amount of a given excited
species which effectively ionise an atom or
molecule, normalised to the abundance of such species and provided 
the energy balance of the process allows for the reaction.
It should be noted, though, that this parameter is unique for
a given gas mixture, i.e. it does not depend on the electric field nor the
high voltage, and that it is expected to vary according to the characteristics
of the quencher(s) and noble gas used in the mixture.
In other words, the Penning transfer can be regarded as a measure of how well
a quencher works: light noble gases tend, 
through their excited states, to ionise quenchers such as CO$_2$,
and therefore the Penning fraction in these mixtures are expected to be
relatively large. On the other hand, heavy noble gases will tend to be 
ionised, probably to a lower extent, by excited molecules of certain
quenchers (Penning mixtures).

\begin{figure}[hbt]
\centering\includegraphics[width=.65\textwidth]{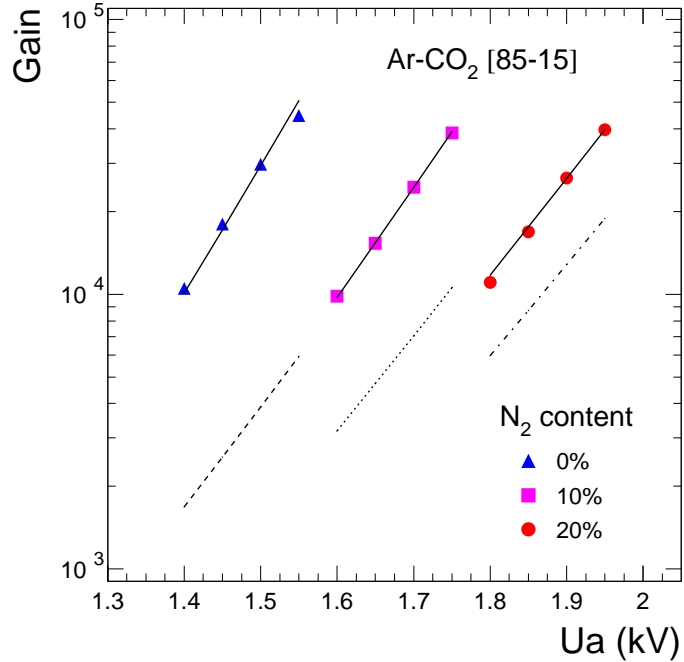}
\caption{Gain in Ar-CO$_2$~[85-15] with N$_2$ additions. The dotted lines are
calculations with Magboltz 2 and no Penning transfer. The tuning of the Penning
transfer parameter to the data yields 37, 20, and 8~\%, for
0, 10, and 20~\% N$_2$, respectively (solid lines).}
\label{aa:arco2g} 
\end{figure} 

In the case of Ar-CO$_2$-N$_2$ mixtures, the suggested Penning mechanism to
provide extra gain is the reaction Ar$^*$ + CO$_2$ $\to$ 
Ar + CO$_2$$^+$ + e$^-$,
where the average excitation energy of the Ar D-levels is 14.0~eV and 
the ionisation potential of CO$_2$ is 13.773~eV.
Fig.~\ref{aa:arco2g} shows the measured and calculated gain as a function 
of anode voltage for the three argon-based mixtures. As can be observed, 
after tuning of the Penning fraction to the second highest point in each curve,
 the slopes are properly matched by the calculations. The Penning fraction 
decreases from 37~\% in the case of no nitrogen to 8~\% when the N$_2$ 
admixture is 20~\%. This means that nitrogen limits the Penning ionisation
of CO$_2$. This effect may occur by quenching of the excited argon 
states by N$_2$ or by the occasional excitation of the nitrogen 
molecule thus leaving the argon atom unexcited. The highest 
excitation level in N$_2$ used in the calculations corresponds to 13.0~eV.
The difference in voltage for equal gain in this series of mixtures is about
200~V, and apparently this gap increases with the gain. 

Shown also in Fig.~\ref{aa:arco2g} are the calculated gains with no Penning
effects, which fail to reproduce the measurements.
In addition, the slopes, at least for the nitrogen free case, 
are less steep than the experimental ones,
and the disagreement of the calculations with the measurements
decreases with increasing N$_2$ concentrations.
Thus, the effect of nitrogen in this mixture, apart from lowering the gain
at a given voltage, is to reduce the Penning effect by providing more
effective quenching.

The case of the Xe-CO$_2$-N$_2$ mixtures is, from the Penning transfer point 
of view, different from argon. In this case, the highest energy level of
excited Xe is 11.7~eV, insufficient to ionise CO$_2$. 
Levels in CO$_2$ between the Xe 
ionisation energy, 12.13~eV, and the CO$_2$ ionisation at 13.773~eV have 
sufficient energy to cause xenon ionisation. Unfortunately, due to the
lack of data, all CO$_2$ excitations above 10.5~eV have been combined into     
a single level at 10.5~eV \cite{aa:bulos,aa:naka} in the simulation program. 
This does not
exclude an analysis similar to the previous mixture since only a fraction 
of the excitation of the 10.5~eV level representing levels above 12.13~eV 
are used in the simulation. In conclusion we assume that the Penning 
transfer occurs from CO$_2^*$(10.5) onto ionisation of xenon. The effect of 
N$_2$ on the Xe-CO$_2$ mixture is quite complex. There are possible   
energy transfer channels from CO$_2^*$ to N$_2$ as in the Ar-CO$_2$ 
mixture but also from N$_2^*$ to ionisation of Xe. The nitrogen excited 
states are produced less copiously than the CO$_2$ excited 
states according to calculations done with
Magboltz 2. Therefore as an approximation we assume the 
dominating transfer is from CO$_2^*$ to Xe.

\begin{figure}[hbt]
\centering\includegraphics[width=.65\textwidth]{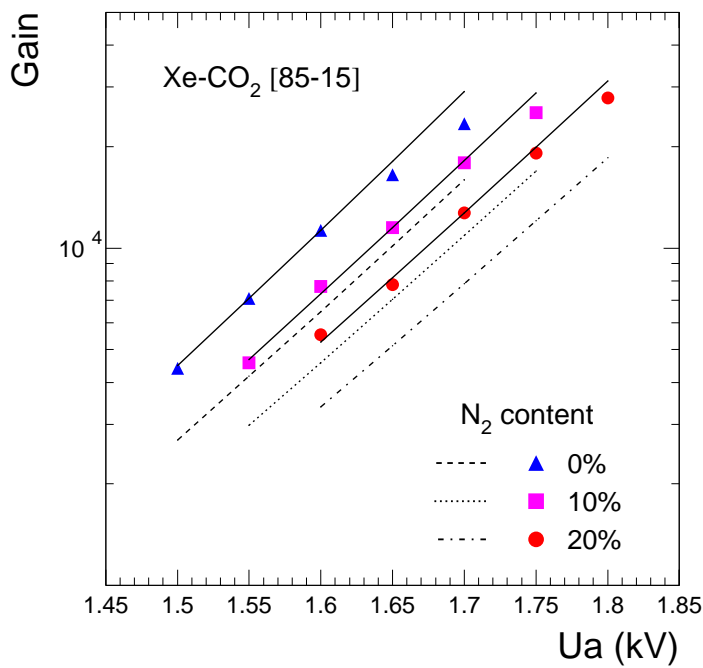}
\caption{Gain in Xe-CO$_2$~[85-15] with N$_2$ additions. The tuned 
Penning transfer rates are 24~\% for the N$_2$-free mixture, 
and 22~\% for the mixtures with N$_2$.}
\label{aa:xeco2g} 
\end{figure}

The experimental gain measurements, and the calculations performed
under these assumptions, shown
in Fig.~\ref{aa:xeco2g}, 
give an approximately constant Penning fraction (22~\%).
All slopes are correctly reproduced, with and without Penning transfer. The
voltage gaps between the curves is about 50~Volts.
The deviation of the data from the calculation -tuned at the middle 
point of each curve- at high gains is probably an indication of space 
charge effects within
the amplification region due to the high X-ray rates. It is interesting
to note that this measured deviation from exponential behaviour seems to 
decrease with increasing N$_2$ concentration, probably due to the higher
anode fields involved. This also implicates space charge as the cause.

\section{Conclusions} \label{aa:sum} 
Drift velocity and gain measurements have been performed for a number of 
gas mixtures in order to assess the effect of nitrogen 
admixture in the gas.
In particular, the drift velocity measurements presented in this work
show a reasonable agreement with calculations performed with Magboltz,
although significant discrepancies are clearly visible in some
cases. Our measurements have been corrected for the effect
of the amplification field leaking between the cathode wires. The effect
of the multiple scattering of sub-MeV electrons in xenon has been reduced 
to a negligible level. In the case of Xe-CO$_2$ mixtures,
the variation of the drift velocity as a function of the N$_2$
admixture turns out to practically vanish at fields below 800~V/cm.

Gain measurements have been performed with mixtures with CO$_2$ and
admixtures of N$_2$. A phenomenological quantification of the Penning
mechanism, namely further ionisation from excited species formed in
the avalanche, has been proposed and calculated with the Magboltz 2 
simulation program. The measured gain curves are not reproduced by the
calculations without this mechanism.
Penning transfer is somewhat inhibited by the presence
of N$_2$ in the argon-based mixtures. In the case of the heavier xenon
mixtures, the role of N$_2$ in this respect seems to be negligible.

\section*{Acknowledgments}
We would like to acknowledge A.~Radu and and J.~Hehner for their skills and 
dedication in building our detectors, and A.~Manafov for his help in
software issues.

\end{document}